\documentclass[12pt,preprint]{revtex4}

\def\de{\delta}
\def\ep{\epsilon}

\def\si{\sigma}

\def\vp{\varphi}

\def\mn{{\mu\nu}}

\def\fr#1#2{{{#1} \over {#2}}}
\def\frac#1#2{\textstyle{{{#1} \over {#2}}}}

\def\lsim{\mathrel{\rlap{\lower4pt\hbox{\hskip1pt$\sim$}}
    \raise1pt\hbox{$<$}}}
\def\gsim{\mathrel{\rlap{\lower4pt\hbox{\hskip1pt$\sim$}}
    \raise1pt\hbox{$>$}}}

\newcommand{\beq}{\begin{equation}}
\newcommand{\eeq}{\end{equation}}
\newcommand{\bea}{\begin{eqnarray}}
\newcommand{\eea}{\end{eqnarray}}
\newcommand{\bse}{\begin{subequations}}
\newcommand{\ese}{\end{subequations}}

\def\nub{\bar\nu}
\def\vp{\vec p}
\def\cmat{{\cal C}}

\def\heff{h_{\rm eff}}

\def\mt{\widetilde m^2}

\begin{document}

\title{Lorentz violation and
  neutrino oscillations\footnote{
    Poster presented at the 22nd International
    Conference on Neutrino Physics and Astrophysics
    (Neutrino 2006), Santa Fe, New Mexico, June 13-19, 2006}}
\author{Matthew Mewes}
\address{Physics Department, Marquette University, Milwaukee, WI 53201}
\begin{abstract}
  Lorentz violation naturally leads
  to neutrino oscillations
  and provides
  an alternative mechanism
  that may explain current data.
  In this work, we discuss possible signals
  of Lorentz violation in 
  neutrino-oscillation experiments.
\end{abstract}

\maketitle

General violations of Lorentz invariance
are described by a theoretical
framework known as the
Standard-Model Extension (SME) 
\cite{cpt,ck1,ck2,kost}.
This program has revealed many
observable consequences and 
led to numerous high-precision
tests of this fundamental symmetry.
One prediction in the neutrino
sector is oscillations caused
by Lorentz violations 
rather than neutrino mass.
This alternative mechanism
results in experimental
signatures that distinguish
it for the more conventional
explanation and lead to 
potential tests of Lorentz invariance
using neutrino-oscillation data.
The resulting sensitivities
rival the most precise
tests in any system.
Here we highlight some of the
key results.
A more detailed discussion can
be found in Refs.\ \cite{km1,km2,km3}.

Leading-order
violations of Lorentz symmetry
in neutrinos are described by
the effective hamiltonian,
\bea
(\heff)_{ab}&=&
|\vp|\de_{ab}
\left(\begin{array}{cc}
1 & 0 \\
0 & 1
\end{array}\right)
+\fr{1}{2|\vp|}
\left(\begin{array}{cc}
(\mt)_{ab} & 0 \\
0& (\mt)^*_{ab}
\end{array}\right)
\nonumber \\
&&+\fr{1}{|\vp|}
\left(\begin{array}{cc}
[(a_L)^\mu p_\mu-(c_L)^\mn p_\mu p_\nu]_{ab} &
-i\sqrt{2} p_\mu (\ep_+)_\nu
[(g^{\mn\si}p_\si-H^\mn)\cmat]_{ab} \\
i\sqrt{2} p_\mu (\ep_+)^*_\nu
[(g^{\mn\si}p_\si+H^\mn)\cmat]^*_{ab} &
[-(a_L)^\mu p_\mu-(c_L)^\mn p_\mu p_\nu]^*_{ab}
\end{array}\right) ,
\nonumber
\label{heff}
\eea
acting on a general neutrino-antineutrino
state vector $(\nu_b,\nub_b)^T$.
The mass-squared matrix $\mt$ provides
the usual Lorentz-conserving theory.
The coefficient matrices 
$(a_L)^\mu$, $(c_L)^\mn$,
$g^{\mn\si}$, and $H^\mn$
control the deviation from perfect Lorentz
symmetry.

Many interesting unconventional effects
arise from this Lorentz-violating
hamiltonian \cite{km1}.
One potential signal
of Lorentz violation in
neutrinos is oscillations
with unusual energy dependences.
Normally, oscillation lengths
are inversely proportional to energy,
so that neutrino oscillation lengths
shrink with an increase in energy.
In contrast,
Lorentz violation
can cause oscillation lengths
that remain constant or grow with energy.
This can lead to
\it spectral anomalies \rm
or distortions in the expected
energy spectrum.

Anisotropies are another important
possibility that arises from a breakdown
of rotational symmetry.
These can cause direction-dependent
oscillations and can have 
profound consequences.
For example,
in terrestrial experiments,
direction-dependent oscillations
can lead to sidereal variations
in the observed fluxes as oscillation
probabilities fluctuate due to
a change in propagation direction
resulting from the rotation of the Earth.

Two strategies have been proposed
for searches for Lorentz violation.
The first involves looking for
model-independent features in
the oscillation data that
can only occur if Lorentz symmetry
is violated.
The other strategy involves
comparing data with
simple candidate test models.
Several models have been proposed
that roughly reproduce current
observations and may help resolve the
LSND anomaly \cite{km2,km3,nu1,nu2}.
In either approach,
extreme sensitivities are expected.
Analyses of current data \cite{nu2,nu3,nu4}
suggest that neutrino-oscillation experiments
have the potential to probe Lorentz
symmetry with sensitivities comparable
to the best tests in any sector.

\end{document}